\documentclass[aps,prl,twocolumn,showpacs,superscriptaddress,10pt]{revtex4-1}

\usepackage{graphicx}
\usepackage{amsmath,amssymb,amstext,amsthm}
\usepackage{bbold}
\usepackage{cancel}
\usepackage{color}
\usepackage{verbatim}
\usepackage{booktabs}
\usepackage[caption=false]{subfig}
\usepackage[english]{babel}
\usepackage{url}

\newcommand{\id}{\mathbb{1}}

\usepackage[utf8x]{inputenc}

\begin{document}

\title{Topological semimetals in the SnTe material class: Nodal lines and Weyl points}

\author{Alexander Lau}
\affiliation{Kavli Institute of Nanoscience, Delft University of Technology, P.O. Box 4056, 2600 GA Delft, Netherlands}

\author{Carmine Ortix}
\affiliation{Institute for Theoretical Physics, Center for Extreme Matter and
Emergent Phenomena, Utrecht University, Princetonplein 5, 3584 CC Utrecht,
Netherlands}
\affiliation{Dipartimento di Fisica ``E. R. Caianiello", Universit\'a di Salerno, IT-84084 Fisciano, Italy}

\date{\today}

\begin{abstract}
We theoretically show that IV-VI semiconducting compounds with low-temperature rhombohedral crystal structure represent a new potential platform for topological semimetals. 
By means of minimal $\mathbf{k}\cdot\mathbf{p}$ models we find that the two-step structural symmetry reduction of the high-temperature rocksalt crystal structure, comprising a rhombohedral distortion along the [111] direction followed by a relative shift of the cation and anion sublattices, gives rise to topologically protected Weyl semimetal and nodal line semimetal phases. We derive general expressions for the nodal features and apply our results to SnTe showing explicitly how Weyl points and nodal lines emerge in this system. 
Experimentally, the topological semimetals could potentially be realized in the low-temperature ferroelectric phase of SnTe, GeTe and related alloys.
\end{abstract}

\maketitle

\paragraph{Introduction -- }
The recent discovery of topological semimetals~\cite{WSC12,LZZ14,HXB15,WFF15,LWF15,LXW15,XBA15,XAB15}, the most prominent examples of which are Weyl semimetals (WSMs) and Dirac semimetals~\cite{VaV14,Wit16,Bur16,YaF17,AMV18}, has attracted 
huge
interest in gapless topological phases of matter~\cite{BuB11,ChS14,Bur15,SGW15,BCW16,Bur17,GNM17,JGM17,GMS17,ODY17,Bur18}. Generally speaking, topological semimetals are systems where the conduction and the valence bands have robust crossings in the Brillouin zone (BZ). In WSMs these robust crossings -- the so-called Weyl nodes -- 
are isolated, two-fold degenerate points and 
generically require the absence of either time-reversal or inversion symmetry~\cite{BuB11,AMV18}.
Furthermore, Weyl points represent monopoles of the Berry flux and therefore carry a topological charge~\cite{Wit16,AMV18}.
The topological nature of Weyl nodes 
leads, by the bulk-boundary correspondence, to
the presence of surface Fermi arcs~\cite{WTV11} possibly coexisting with surface Dirac cones~\cite{LKB17}.
In Dirac semimetals, instead, both the conduction and the valence bands are two-fold degenerate and cross at 
isolated four-fold degenerate points
in the BZ. As opposed to Weyl nodes,
Dirac points are 
typically unstable
degeneracies and can be regarded
as the parent semimetallic state generating 
a WSM
by inversion or time-reversal symmetry breaking~\cite{YZT12}.
Various WSM 
materials
have been
predicted theoretically~\cite{Mur07,BuB11,HaB12,Hos12,WTV11,XWW11,BLQ14,WFF15,SGW15,RJY16,LaO17} and realized experimentally~\cite{HXB15,LWF15,LXW15,XBA15,XAB15,XWL16,DWD16,WZH16,KKE16}.
These include both binary 
and ternary compounds~\cite{note1}.
Nevertheless, in view of potential applications, it is important to seek for new material platforms and novel mechanisms for the realization of WSMs. 

A different
class of topological semimetals features conduction and valence bands crossing each other along closed \emph{lines} in the BZ~\cite{BHB11,FCK15,HLV16,FWD16,XYF17}.
These nodal line semimetals 
are
midway between semimetals with point nodes and 
metals with a two-dimensional Fermi surface. 
One of the typical features of nodal line semimetals is the presence of drumhead surface states bounded by the surface projection
of the nodal lines, whose stability is guaranteed by the presence of, for instance, mirror symmetries~\cite{BCZ16,CCC16,HLV16}.
In contrast to WSMs, only few candidate materials for topological nodal line semimetals have been put forward~\cite{FWD16}.

In this Letter, we show that both nodal line and WSM phases can potentially appear as a result of a structural distortion in 
group-IV tellurides
with high-temperature rocksalt crystal structure,
such as SnTe, GeTe and related alloys~\cite{RaJ85,SSM10,MMF17}. A crystal symmetry reduction to a rhombohedral phase via an elastic
strain along the (111) direction 
reduces the point group symmetry of a subset of $L$ points in the BZ~\cite{Fer65}. We show that this 
leads to bulk Dirac points close to these high-symmetry points that 
evolve either into pairs of Weyl nodes or into mirror-symmetry protected nodal loops 
upon breaking inversion symmetry.
The latter is naturally realized, for instance, via a relative shift of the anion and cation sublattices during a ferroelectric distortion.
Our analysis is based on effective $\mathbf{k}\cdot\mathbf{p}$ models describing the low-energy physics close to the $L$ points of the BZ. 
In particular, we derive general expressions for Weyl points and nodal lines, and apply our general results to a specific model based on SnTe.
We show explicitly how Weyl points and nodal lines appear and calculate topological invariants associated with the semimetallic phases.

\paragraph{Dirac points by strain engineering --} 
IV-VI narrow band gap semiconductors have a high-temperature
rocksalt lattice structure with a face-centered cubic 
BZ~\cite{HLL12}.
The BZ is bounded by six square faces and eight hexagonal faces. The centers of the latter, commonly denoted by $L$, represent high-symmetry points in the BZ with $D_{3d}$ point group symmetry~\cite{Fer65}, which is
generated by inversion, a $C_3$ axis along $\Gamma L$
and a mirror plane containing $\Gamma$ and two
$L$ points, hereafter dubbed as $L$ and $L'$, related by a $C_4$ rotation~\cite{HLL12}. 

Since the fundamental band gap of group-IV tellurides
is located at the four equivalent $L$ points related by the point group symmetries of the lattice~\cite{MiW66}, the band structure close to the Fermi level can be captured within an effective four-band low-energy $\mathbf{k}\cdot\mathbf{p}$ model~\cite{HLL12}.
We start out by taking this continuum model and augment it by terms quadratic in the momentum $\mathbf{k}$.
Taking into account all symmetry constraints, 
including time-reversal symmetry,
(see Refs.~\cite{VRA18}, \cite{qsymm}, and Supplemental Material (SM)~\cite{supp})
the model reads:
\begin{eqnarray}
H_0(\mathbf{k}) &=& m\sigma_z + \nu(k_1 s_2 - k_2 s_1) \sigma_x 
+ \nu_3 k_3 \sigma_y \nonumber\\
&&{}+ c k_3^2 \sigma_z + f (k_1^2 + k_2^2)\sigma_z,
\label{eq:full_symmetry_Hamiltonian}
\end{eqnarray}
where, without loss of generality, we have neglected all terms proportional to the identity since they correspond either to a rigid shift of all energies or to a balanced change in the curvature of all bands. 
Therefore, they do not affect the band topology.
In the chosen coordinate system, $k_1$ is perpendicular to the mirror plane, and $k_3$ points along the $C_3$ axis going through the $L$ point. 
The
$\sigma_i$ are Pauli matrices in orbital space spanned by
the $p$ orbitals of
the cation (Pb,~Sn,~Ge) and anion (Te),
whereas the $s_i$ are Pauli matrices in spin space. 
Due to the simultaneous presence of inversion and time-reversal symmetry, all states are two-fold degenerate.

Contrary to the trivial semiconductors PbTe and GeTe, for
SnTe
it is well known that an inverted band gap at the $L$ points gives rise to a crystalline topological-insulating phase protected by mirror symmetry~\cite{HLL12}. Now, 
we show that, \emph{independent} of the band ordering, a structural distortion to a rhombohedral phase via an elastic strain along the cube diagonal, {\it i.e.}, the [111] direction, can
lead to the emergence of
bulk Dirac points, i.e., generic four-fold degenerate band crossing points.  

\begin{figure}[t]\centering
\includegraphics[width=1.0\columnwidth]
{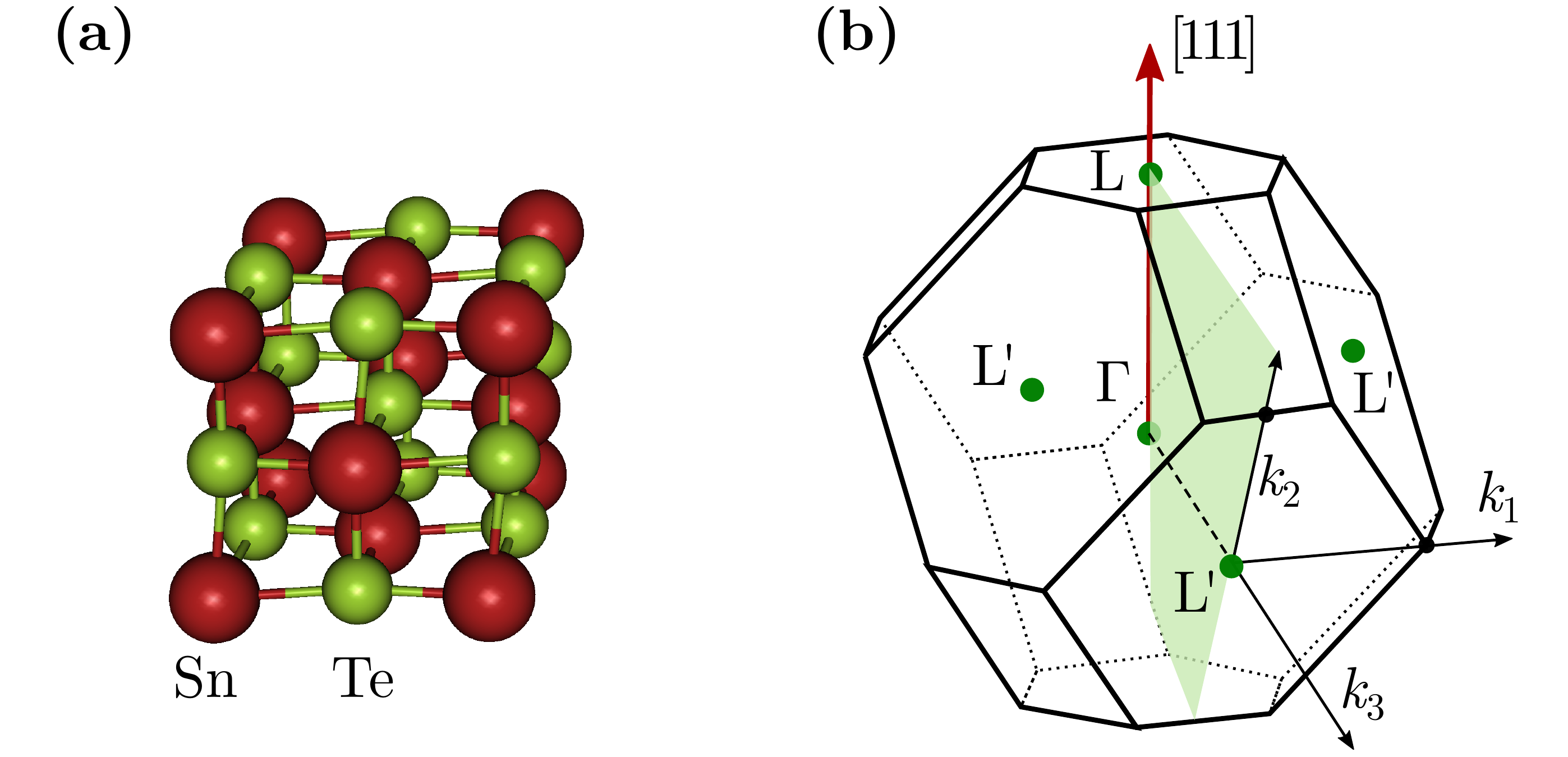}
\caption{(a) Crystal structure of rhombohedrally distorted SnTe. (b) Rhombohedral Brillouin zone with high-symmetry points $L$ and $L'$, a local coordinate system at one of the $L'$ points, and the strain direction (red arrow) responsible for the distortion from a face-centered cubic to a rhombohedral lattice. Also highlighted is one of the three mirror planes spanned by $\Gamma$, $L$ and $L'$.}
\label{fig:lattice}
\end{figure}

The rhombohedral distortion breaks the $C_4$ symmetry of the face-centered cubic lattice [see Fig.~\ref{fig:lattice}(a)]. Consequently, 
the square faces of the BZ distort into rectangles and the hexagonal faces are no longer identical as is illustrated in Fig.~\ref{fig:lattice}(b).
Equivalently, the corresponding elastic strain
acts differently on different $L$ points~\cite{Fer65}:
it does not affect the symmetry of the $L$ point in the [111] direction. However, in the local coordinate system of the point $L'$, 
which was previously related to $L$ by a $C_4$ rotation,
the strain acts in the $[11\bar{1}]$ direction. In contrast to $L$, this lowers the symmetry group at $L'$ from $D_{3d}$ to $C_{2h}$:
the $C_3$ symmetry with respect to an axis going through $L'$ is explicitly broken, whereas mirror and inversion symmetry are still preserved. Equivalently, the symmetry group at the other two 
$L'$ points is lowered to $C_{2h}$. 
As a result,
there are now one $L$ point with $D_{3d}$ symmetry and three $L'$ points with $C_{2h}$ symmetry.

The reduction of point-group symmetry at the $L'$ points gives rise to
additional symmetry-allowed terms in the corresponding
$\mathbf{k}\cdot\mathbf{p}$ theory. Up to second order in $\mathbf{k}$ they read
\begin{eqnarray}
H_1(\mathbf{k}) &=& \delta\nu (k_1 s_2 + k_2 s_1)\sigma_x + \lambda_1 k_1 s_3 \sigma_x 
+ \lambda_2 k_2 \sigma_y \nonumber\\
&&{}+ \lambda_3 k_3 s_1 \sigma_x + \delta f (k_1^2 - k_2^2)\sigma_z + g k_2 k_3 \sigma_z.
\label{eq:symmetry_breaking_terms}
\end{eqnarray}
We now show
that the extended model given by $\tilde{H}=H_0+H_1$ gives rise to isolated bulk Dirac points. 
The energies of $\tilde{H}$ 
can be written as
\begin{eqnarray}
E^2 &=& (f_1 k_1^2 + f_2 k_2^2 + g k_2 k_3 + c k_3^2 + m)^2\nonumber\\ 
&&{}+ k_1^2 (\lambda_1^2 + \nu_1^2) + (k_2 \lambda_2 + k_3 \nu_3)^2\nonumber\\
&&{}+ (k_3 \lambda_3 - k_2 \nu_2)^2,
\end{eqnarray}
where we have defined $\nu_{1,2}=\nu\pm\delta\nu$ and $f_{1,2}=f \pm \delta f$. The spectrum is symmetric under $E\rightarrow -E$. Hence, Dirac points, if present, will be located at $E=0$. It is straight forward to see that the spectrum has Dirac points if $E$ is of the form $\pm\sqrt{(ak^2 - m)^2}$. This implies that all binomials under the square root have to vanish, except the first. From this we determine the single condition $\lambda_2\lambda_3 = -\nu_2\nu_3$,
which can be satisfied by tuning the external strain magnitude. Under this condition, Dirac points, if present, will be located in the mirror plane on the line parametrized by $k_3=\nu_2 k_2 / \lambda_3$ and $k_1=0$.

To focus only on the essential mechanism leading to the existence of Dirac points, we will neglect all terms in $H_1$ that do not enter the conditions above explicitly, i.e., we set $g=\lambda_1=0$, $f_1=f_2=f$, and $\nu_1=\nu_2=\nu$. 
We emphasize that this is done merely 
to simplify
our analytical considerations.
The results presented below can, however, be generalized also to the full model.
With this and the constrains given above, the spectrum along the line $\mathbf{k} = (0, k, \nu k / \lambda_3)$ becomes
\begin{equation}
E_\pm(k) = \sqrt{\bigg[\bigg(\frac{c\nu^2}{\lambda_3^2} + f \bigg)\, k^2 + m \bigg]^2}.
\label{eq:Dirac_spectrum}
\end{equation}
This has the desired form and we infer: Dirac points exist if $m$ and the term before $k^2$ have \emph{opposite} sign.
This can be realized by tuning the band mass $m$, \textit{e.g.} by alloying or pressure~\cite{BRS13}.
The Dirac points are located at $\pm\mathbf{k}_0$ with $(k_{0,1}, k_{0,2}, k_{0,3}) = (0, \sqrt{-m\lambda_3/(c\nu^2 + f\lambda_3^2)}, \nu/\lambda_3\, k_{0,2})$. Moreover, an expansion of $\tilde{H}$ around the Dirac points to
leading order
in $\mathbf{k}$ shows that the effective Hamiltonian is indeed of Dirac form (see SM~\cite{supp}) 
with, in general, anisotropic dispersion.

\begin{figure}[t]\centering
\includegraphics[width=1.0\columnwidth]
{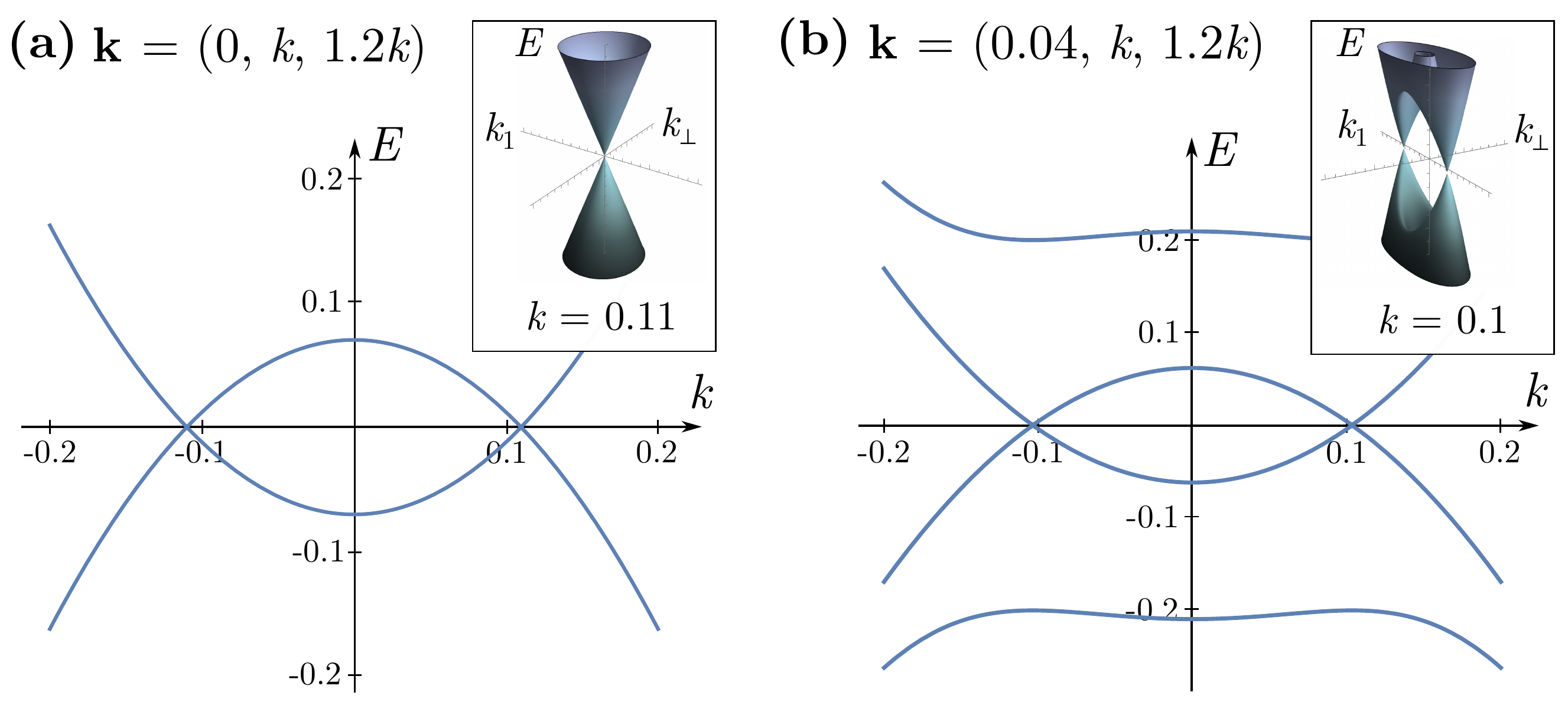}
\caption{Spectra of SnTe models with $\lambda_2=-1.14\,\mathrm{eV}$ and $\lambda_3=2.0\,\mathrm{eV}$ along a line through the local momentum-space coordinate system centered at $L'$. (a) With inversion symmetry: the two bands are each two-fold degenerate and cross at two isolated Dirac points. (b) Without inversion symmetry and $\alpha=0.1\,\mathrm{eV}$: the spectrum features four separate Weyl points. The insets show spectra along a perpendicular plane in $\mathbf{k}$ space going through one of the band crossings in the associated main panel. Energies $E$ are given in [eV]. Momenta $\mathbf{k}$ are displayed in units of [$\pi/a$] with the lattice constant $a$.}
\label{fig:Dirac_and_Weyl_points}
\end{figure}

We
now apply our general results to a specific system by means of numerical calculations. 
To
obtain realistic values for the $\mathbf{k}\cdot\mathbf{p}$ parameters of our model, we fit the parameters of the Hamiltonian $H_0$ in Eq.~\eqref{eq:full_symmetry_Hamiltonian} to 
density functional theory
data of SnTe presented in Ref.~\cite{HLL12}. From that, we determine the following values (in eV): $m=-0.07$, $\nu=2.4$, $\nu_3=0.95$, $c=0.9$, and $f=4.5$. Next, we introduce a rhombohedral distortion in our SnTe model by tuning $\lambda_2$ and $\lambda_3$ away from zero until the Dirac-point condition $\lambda_2=-\nu\nu_3/\lambda_3$ is established. The resulting spectrum along a cut through the local coordinate system is shown in Fig.~\ref{fig:Dirac_and_Weyl_points}(a). We find two Dirac points in agreement with the analytical prediction.

\paragraph{Weyl points and nodal lines --}
Bulk Dirac points are, in general, unstable features and can be gapped out by small perturbations. Nonetheless, it is well-known that a bulk Dirac point can be split into a pair of stable Weyl points of opposite charge 
by breaking inversion symmetry.
Furthermore, 
if mirror symmetry is still present,
a Dirac point can also evolve into a nodal line protected by this symmetry~\cite{SBZ18}. 
These conditions are naturally realized in the ferroelectric phase of SnTe and GeTe: below a critical temperature~\cite{YDW15} $T_c=98$~K ($T_c=670$~K), SnTe (GeTe) undergoes
a structural transition from a rocksalt structure with space group $Fm\bar{3}m$ to a rhombohedral lattice with space group $R3m$~\cite{RaJ85,SSM10}. This transition occurs via a two-step symmetry reduction~\cite{RaJ85,SSM10,MMF17}. First, an elastic strain along the cube diagonal introduces a rhombohedral distortion and breaks the $C_4$ symmetry. This is identical to the symmetry reduction process discussed above. Second, a relative displacement of the Sn (Ge) and Te sublattices breaks spatial inversion symmetry, a necessary condition for the semimetallic phases considered in this work, while preserving mirror and $C_3$ symmetries. The distorted lattice is illustrated in Fig.~\ref{fig:lattice}(a).

To incorporate the second step of the symmetry-reduction procedure into our model, we note that inversion-symmetry breaking reduces the symmetry group of the $L'$ points further from $C_{2h}$ to $C_s$, i.e., only the mirror plane remains. In total, there are 10 additional symmetry-allowed terms (see SM~\cite{supp}). Here, we restrict our consideration to the following terms
\begin{eqnarray}
H_\alpha(\mathbf{k}) &=& \alpha\sigma_x,
\label{eq:inversion_breaking_alpha}\\
H_\beta(\mathbf{k}) &=& \beta(k_1 s_2 - k_2 s_1),
\label{eq:inversion_breaking_beta}
\end{eqnarray}
because each of them gives rise to one of the nodal features described above 
in a straight-forward fashion. We note, however, that also the other inversion-symmetry breaking terms give rise to the same features.

We 
next
show that the term $H_\alpha$ in Eq.~\eqref{eq:inversion_breaking_alpha} splits the Dirac point into two stable Weyl points. For that, we expand the Hamiltonian $\tilde{H}+H_\alpha$ around the Dirac point at $\mathbf{k}_0$ up to leading order in momentum. The effective Hamiltonian (see SM~\cite{supp}) has the following spectrum,
\begin{eqnarray}
E^2 &=& (2 f k_{0,2} \kappa_2 + 2 c k_{0,3} \kappa_3)^2 
+ \frac{\nu_3^2}{\lambda_3^2} (\lambda_3 \kappa_3 - \nu\kappa_2)^2 \nonumber\\
&&{}+ \Big(\alpha \pm \sqrt{(\lambda_3 \kappa_3 - \nu\kappa_2)^2 + (\nu\kappa_1)^2}\Big)^2.
\label{eq:energies_alpha}
\end{eqnarray}
Since we are again looking for zero-energy states, all terms in parentheses in the equation above must simultaneously vanish. We already know that this is the case for $\alpha=0$. Keeping all parameters fixed except $\alpha$, this implies that zero-energy states must satisfy $\kappa_2=\kappa_3=0$ even for nonzero $\alpha$. Finally, we obtain zero-energy solutions of Eq.~\eqref{eq:energies_alpha} for
\begin{equation}
\mathbf{k}_W = (\pm \alpha/\nu, k_{0,2}, k_{0,3}).
\label{eq:Weyl_point_position}
\end{equation}
The solutions are distinct for $\alpha\neq 0$ and each of them is two-fold degenerate. Furthermore, the Weyl points are mapped onto each other by reflection about the mirror plane. Since reflection flips the topological charge of a Weyl node~\cite{SBZ18}, we further infer that their topological charge must be opposite.
These general findings are confirmed by numerical results as we show in Fig.~\ref{fig:Dirac_and_Weyl_points}(b): the two Dirac points split into four two-fold degenerate states, two on each side of the mirror plane. Furthermore, we calculate the topological charge of each nodal point numerically~\cite{FHS05} and obtain nontrivial values of $\pm 1$. 

We emphasize that, due to their topological charge, the Weyl nodes are robust features of the system and, thus, must appear in an extended region in the parameter space. This implies that we can now explicitly violate the condition $\lambda_2\lambda_3 = -\nu_2\nu_3$, which led to the existence of Dirac points, or switch on other parameters without gapping out the Weyl nodes (see SM~\cite{supp}). Hence, the Weyl nodes are not subject to parameter fine-tuning, which is in stark contrast to the parent Dirac points.

\begin{figure}[t]\centering
\includegraphics[width=1.0\columnwidth]
{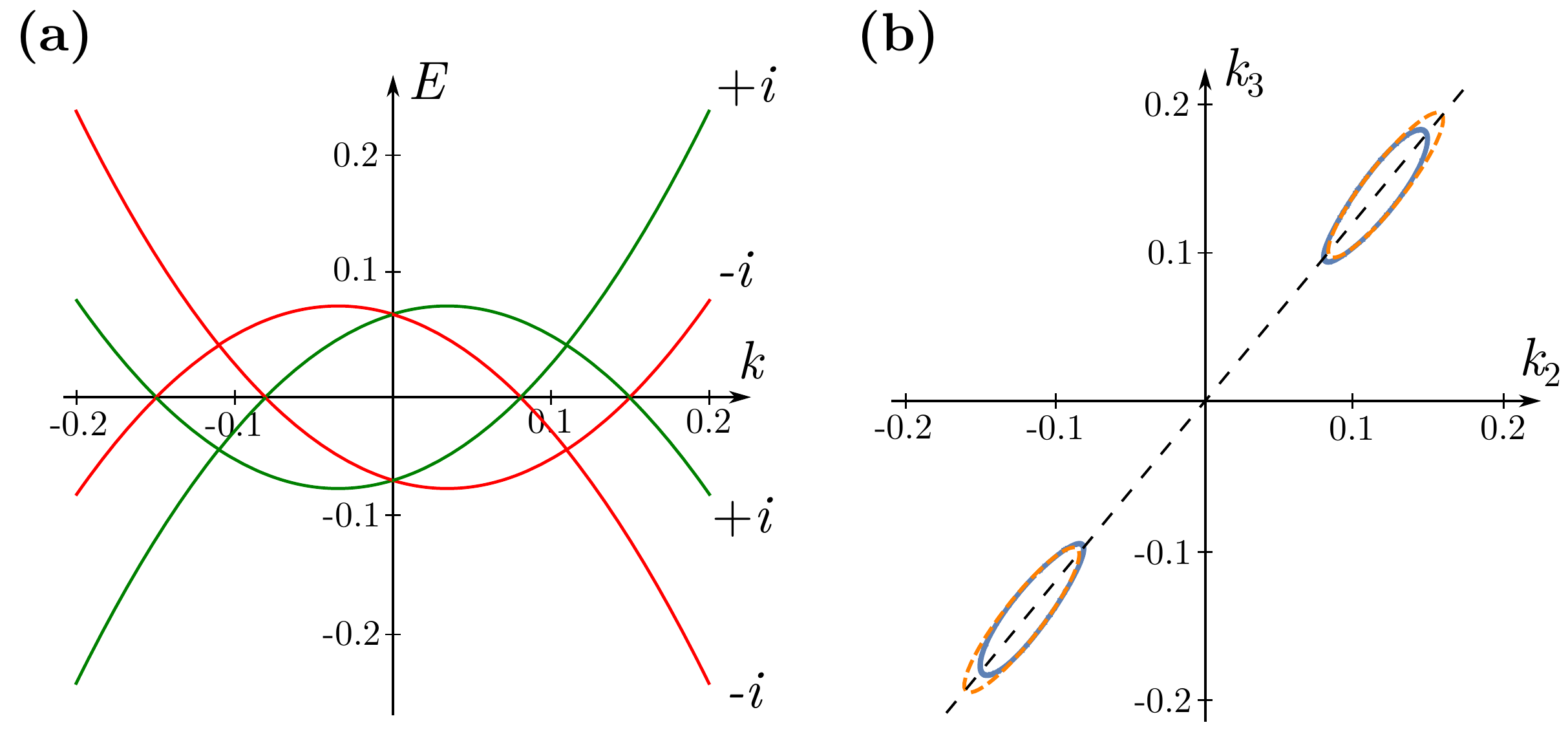}
\caption{Spectrum of the SnTe model with $\lambda_2=-1.14\,\mathrm{eV}$, $\lambda_3=2.0\,\mathrm{eV}$ and $\beta=0.4\,\mathrm{eV}$ (broken inversion symmetry) in the local momentum-space coordinate system centered at $L'$. (a) Dispersion along a line parametrized by $\mathbf{k}=(0,k,1.2k)$: there are four two-fold degenerate band crossings at zero energy. The mirror eigenvalues of the bands have been indicated in green ($+\mathrm{i}$) and red ($-\mathrm{i}$). (b) $E=0$ Fermi surface (blue lines) in the mirror plane at $k_1=0$. The spectrum features two nodal loops. For comparison, the approximate analytical solutions for the nodal lines have been indicated (dashed orange lines). The black dashed line represents the momentum-space cut shown in (a). Energies $E$ are given in [eV]. Momenta $\mathbf{k}$ are displayed in units of [$\pi/a$].}
\label{fig:nodal_lines}
\end{figure}

Due to the presence of mirror symmetry, the decay of a Dirac point into two Weyl points is not the only possible process. In fact, 
a Dirac point can also evolve into a topologically protected nodal loop located in the mirror plane~\cite{SBZ18}. 
We now
show that this is exactly what happens 
using
$H_\beta$ from Eq.~\eqref{eq:inversion_breaking_beta} to break inversion symmetry.

As before,
we first expand the Hamiltonian $\tilde{H}+H_\beta$ around the Dirac point at $\mathbf{k}_0$ (see SM~\cite{supp} for the resulting effective Hamiltonian). Let us look at this Hamiltonian along the $k_1$ direction perpendicular to the mirror plane. The spectrum along this line is
\begin{equation}
E(\kappa_1,0,0) = \pm\sqrt{(\beta k_{0,2})^2 + \kappa_1^2 (\beta\pm\nu)^2},
\label{eq:energies_beta_k1}
\end{equation}
which is always nonzero for $\kappa_1\neq 0$, even for infinitesimally small values of $\beta$. Zero-energy states are therefore expected to
be located in the mirror plane at $k_1=\kappa_1=0$.

To obtain the energies in the mirror plane, we look at the spectrum along arbitrary lines through the Dirac point. We parametrize these lines by their slope $\eta$, i.e., we look at lines of the form $(0, \kappa, \eta\kappa)$. We obtain
\begin{eqnarray}
E_\eta(\kappa) &=& \pm (\kappa + k_{0,2}) \pm \frac{\sqrt{A_\eta}}{\lambda_3}\,\kappa,
\end{eqnarray}
where 
$A_\eta = 4\lambda_3^2 (f k_{0,2} + c k_{0,3}\eta)^2 
+ (\nu - \lambda_3\eta)^2 (\lambda_3^2 + \nu_3^2)$. From this, we determine the location of zero-energy states 
\begin{eqnarray}
\mathbf{k}_{N,\eta} &=& \frac{-\lambda_3\beta k_{0,2}}{\lambda_3\beta \pm \sqrt{A_\eta}}\,
(0,1,\eta).
\label{eq:nodal_lines_position}
\end{eqnarray}
Each state is two-fold degenerate similar to the Weyl-point solutions above. However, the structure of the solutions is qualitatively different
as we show in the SM~\cite{supp}: the set of zero-energy states form a closed line.

We again check our analytical findings against numerical results for our SnTe model. The spectrum along a cut through the BZ is shown in Fig.~\ref{fig:nodal_lines} along with the $E=0$ Fermi surface in the mirror plane $k_1=0$. In accordance with our analytical study, 
the Dirac points evolve into elliptical two-fold degenerate lines located in the mirror plane. 

The nodal lines are topologically protected by mirror symmetry~\cite{SBZ18,FWD16}. 
The
mirror plane at $k_1=0$, accommodating the nodal lines, is invariant under reflection. Therefore, the mirror operator $M$ and the system Hamiltonian $H(\mathbf{k})$ commute for all momenta in this plane and all states can be assigned a well-defined reflection eigenvalue $\xi=\pm i$. This is illustrated with different colors in Fig.~\ref{fig:nodal_lines}(a). We observe that the reflection eigenvalues of occupied states inside the nodal ellipses are different from the ones outside the ellipses, namely, $\xi_\mathrm{in} = \lbrace \mp i,\:\mp i\rbrace$ and $\xi_\mathrm{out} = \lbrace \mp i,\: \pm i\rbrace$ (for $k\gtrless 0$). These values cannot change unless the bulk energy gap closes locally, which happens along the nodal lines. Therefore, the nodal lines are protected. The corresponding topological invariant is computed from the difference of occupied states with mirror eigenvalue $-i$ inside and outside the ellipses~\cite{FWD16}. We obtain $\nu_\mathrm{line}=\pm 1$ for $k\gtrless 0$. 

\paragraph{Conclusions --}

We have shown that systems in the SnTe material class are a new potential platform for Weyl and nodal line semimetals. 
The key ingredients are a rhombohedral distortion, induced by an elastic strain along the [111] direction, followed by inversion-symmetry breaking. By deriving and analyzing a minimal $\mathbf{k}\cdot\mathbf{p}$ model, we have demonstrated how this two-step symmetry reduction process leads to the appearance of topologically stable Weyl nodes. Complementary, 
we find that the mirror symmetry in group-IV tellurides also gives rise to a semimetallic phase featuring topologically protected nodal lines, a class of systems for which only few candidate materials have been 
put forward.

There are various, feasible ways to 
realize our proposal experimentally.
First,
the conditions for the symmetry reduction are naturally provided in the low-temperature ferroelectric phase of SnTe and GeTe, which could therefore represent novel Weyl ferroelectric semimetals~\cite{SBS16, LXH16}.
Moreover, additional external strain, pressure, or alloying~\cite{DMS66,Str67,SSM10,JCB13,MMF17,LBS18} 
could be employed to tune the parameters of the systems. In fact, a recent experimental report
supports the existence of semimetallic phases in Pb-alloyed SnTe under pressure~\cite{LKK17}.
Another
route is the use of substrates with different lattice structures. This could induce an inhomogeneous strain close to the substrate interface
mimicking a ferroelectric distorsion.

Finally, our proposal could also be applied to 
group-V 
semimetals
such as Bi and Sb~\cite{LiA95}. These materials are in the rhombohedral space group $R{\bar 3}m$ with inversion center and have a similar band structure as group-IV tellurides~\cite{GoG74}. To realize Weyl nodes or nodal lines, one could therefore
use thin films where inversion symmetry can be broken by either using substrates, as explained above, or externally applying a perpendicular electric field.

\begin{acknowledgments}
We thank D. Varjas and T. Rosdahl for helpful discussions regarding the symmetric-Hamiltonian generator algorithm. 
C.O. acknowledges support from 
a VIDI grant (Project 680-47-543) financed by the Netherlands Organization for Scientific Research (NWO).
A.L. acknowledges support from the Netherlands Organisation for Scientific Research (NWO/OCW), as part of the Frontiers of Nanoscience program.
\end{acknowledgments}


\begin{thebibliography}{68}
\bibitem [{not()}]{note1}%
  \BibitemOpen
  \href@noop {} {}\bibinfo {note} {For a more detailed and exhaustive overview
  of the field, we refer to the excellent review article in
  Ref.~\cite{AMV18}.}\BibitemShut {Stop}%
\end{thebibliography}

%


\appendix

\section*{SUPPLEMENTAL MATERIAL}

\section*{A: Derivation of the extended $\mathbf{k}\cdot\mathbf{p}$ model}

A central part of our study is the derivation of a minimal $\mathbf{k}\cdot\mathbf{p}$ model, based on the 4-band model in Ref.~\cite{HLL12}, for systems in the SnTe material class taking into account all symmetry-allowed terms up to second order in $\mathbf{k}$. For this purpose, we have used 
an algorithm,
provided in the Python package Qsymm~\cite{VRA18,qsymm}, that systematically generates all possible terms of a Hamiltonian up to a given order in $\mathbf{k}$ respecting a given set of symmetries. In the following, we are going to provide all the ingredients required for the application of the symmetry algorithm.

Our model has two orbital degrees of freedom, 
spanned by $p$ orbitals on Sn and Te sites, represented by Pauli matrices $\sigma_i$, and two spin degrees of freedom represented by Pauli matrices $s_i$~\cite{HLL12}. The $\mathbf{k}\cdot\mathbf{p}$ model is derived with respect to an $L$ point of the, initially, face-centered cubic BZ. The initial symmetry group of the $L$ point is $D_{3d}$ which is generated by inversion $I$, a rotation $R_3$ about the $C_3$ axis along $\Gamma L$, and a reflection $M$ about the mirror plane containing $\Gamma$ and two $L$ points. Furthermore, the model should be invariant under time reversal $\Theta$. The corresponding representations of the symmetry operators are as follows,
\begin{eqnarray}
M &=& -i\,s_1,\:\: k_1\rightarrow -k_1,\\
R_3 &=& e^{i\frac{\varphi}{2}s_3},\:\mathbf{k}\rightarrow\!\!
\begin{pmatrix}
\cos\varphi & -\sin\varphi & 0\\
\sin\varphi & \cos\varphi & 0\\
0 & 0 & 1
\end{pmatrix}\! \mathbf{k},\: \varphi=\frac{2\pi}{3}\\
I &=& \sigma_z,\:\: \mathbf{k}\rightarrow -\mathbf{k},\\
\Theta &=& is_2\,K,\:\: \mathbf{k}\rightarrow -\mathbf{k},
\end{eqnarray}
where $k_i$ are momentum components with respect to a local coordinate system at $L$ spanned by $\hat{\mathbf{k}}_1$ perpendicular to the mirror plane, $\hat{\mathbf{k}}_3$ pointing along the $C_3$ axis, and $\hat{\mathbf{k}}_2$ such that $\lbrace \hat{\mathbf{k}}_1,\hat{\mathbf{k}}_2,\hat{\mathbf{k}}_2\rbrace$ form a right-handed coordinate system. 
The $\sigma_z$ in the inversion operator is a result of expanding around an $L$ point: because the inversion center is located at one of the lattice sites in the unit cell of the rocksalt structure, the other site is translated by a lattice vector under inversion and, thus, acquires a phase factor at nonzero momentum.

By providing \emph{all} operators above as input, we apply the $\mathbf{k}\cdot{\mathbf{p}}$ Hamiltonian generator algorithm of the Qsymm package~\cite{VRA18,qsymm} and find 8 symmetry-allowed terms. Ignoring the 3 terms that are proportional to the identity and do not influence the band topology, we obtain the following Hamiltonian
\begin{eqnarray}
H_0(\mathbf{k}) &=& m\sigma_z + \nu(k_1 s_2 - k_2 s_1) \sigma_x 
+ \nu_3 k_3 \sigma_y \nonumber\\
&&{}+ c k_3^2 \sigma_z + f (k_1^2 + k_2^2)\sigma_z,
\label{eq:full_symmetry_Hamiltonian_supp}
\end{eqnarray}
We proceed with the first step of symmetry reduction process during which the $C_3$ symmetry of the model is broken. By repeating the Hamiltonian generator algorithm with only the symmetry operators $M$, $I$, and $\Theta$, 
we find 8 additional terms, 6 of which are not proportional to the identity:
\begin{eqnarray}
H_1(\mathbf{k}) &=& \delta\nu (k_1 s_2 + k_2 s_1)\sigma_x + \lambda_1 k_1 s_3 \sigma_x 
+ \lambda_2 k_2 \sigma_y \nonumber\\
&&{}+ \lambda_3 k_3 s_1 \sigma_x + \delta f (k_1^2 - k_2^2)\sigma_z + g k_2 k_3 \sigma_z.
\label{eq:symmetry_breaking_terms_supp}
\end{eqnarray}
Finally, breaking inversion symmetry is incorporated by repeating the algorithm with only the symmetry operators $M$ and $\Theta$. This leads to 10 additional symmetry-allowed terms up to leading order in $\mathbf{k}$:
\begin{eqnarray}
H_2(\mathbf{k}) &=& \alpha \sigma_x + \beta(k_1s_2-k_2s_1) 
+ \delta\beta(k_1s_2 + k_2s_1)\nonumber\\ 
&&{} + \gamma(k_1s_2 - k_2s_1)\sigma_z + \delta\gamma(k_1s_2 + k_2s_1)\sigma_z \nonumber\\
&&{}+ \eta_1 k_1s_z + \eta_2 k_1s_z\sigma_z + \eta_3 k_3s_x \nonumber\\
&&{}+ \eta_4 k_3s_x\sigma_z + \eta_5 s_x\sigma_y
\end{eqnarray}
In the main text we have shown that the terms parametrized by $\alpha$ and $\beta$ give rise to Weyl points and nodal lines. Moreover, it can be checked straightforwardly that also the other terms give rise to the same features.

\section*{B: Effective Hamiltonians around the Dirac point}

In the main text we derive a condition on the existence of Dirac points in the $\mathbf{k}\cdot\mathbf{p}$ Hamiltonian given by
\begin{eqnarray}
H(\mathbf{k}) &=& m\sigma_z + \nu(k_1 s_2 - k_2 s_1) \sigma_x 
+ \nu_3 k_3 \sigma_y\nonumber\\
&&{}+ c k_3^2 \sigma_z + f (k_1^2 + k_2^2)\sigma_z  \nonumber\\
&&{}+ \lambda_2 k_2 \sigma_y
+ \lambda_3 k_3 s_1 \sigma_x.
\label{eq:kp_Hamiltonian}
\end{eqnarray}
In particular, under the condition $\lambda_2\lambda_3 = -\nu\nu_3$ there exist isolated, four-fold degenerate zero-energy states at
\begin{eqnarray}
\mathbf{k}_0 &=& 
(k_{0,1}, k_{0,2}, k_{0,3}) \nonumber\\
&=& 
(0, \sqrt{-m\lambda_3/(c\nu^2 + f\lambda_3^2)}, \nu/\lambda_3\, k_{0,2}).
\end{eqnarray}
In the following, we are going to have a closer look at the structure of the Hamiltonian close to $\mathbf{k}_0$.

\subsection*{B.1 Dirac Hamiltonian}

Let us expand the Hamiltonian of Eq.~\eqref{eq:kp_Hamiltonian} around $\mathbf{k}_0$ up to leading order in the momentum $\kappa=\mathbf{k}-\mathbf{k}_0$. The resulting effective Hamiltonian is
\begin{eqnarray}
H_\mathrm{eff}(\kappa) &=& (2c \frac{\nu k_{0,2}}{\lambda_3}\sigma_z + \lambda_3 s_1 \sigma_x + \nu_3 \sigma_y)\, \kappa_3\nonumber\\
&&{} + (2f k_{0,2}\sigma_z - \nu s_1\sigma_x - \frac{\nu\nu_3}{\lambda_3}\sigma_y)\, \kappa_2\nonumber\\
&&{} +\nu s_2\sigma_x \kappa_1.
\label{eq:effective_Dirac_Hamiltonian}
\end{eqnarray}
The effective Hamiltonian has no zeroth order terms and is indeed linear. Moreover, the involved matrices 
$\gamma_0= \sigma_y\otimes\id$, 
$\gamma_1= \sigma_x\otimes s_2$, 
$\gamma_2= \sigma_z\otimes\id$, 
$\gamma_3= \sigma_x\otimes s_1$
satisfy $\lbrace \gamma_i, \gamma_j\rbrace = 0$ for $i \neq j$ and $\gamma_i^2=\id$. Therefore, they form a Clifford algebra. We can further form the chiral operator $\gamma_5=\gamma_0\gamma_1\gamma_2\gamma_3=\sigma_x\otimes s_3$ which satisfies $\gamma_5^2=\id$ and which anticommutes with the Hamiltonian, i.e.,  $\gamma_5 H_\mathrm{eff}(\kappa)\gamma_5 = -H_\mathrm{eff}(\kappa)$. Thus, the effective Hamiltonian has a chiral symmetry. 

By defining new $\gamma$ matrices through linear combinations of $\gamma_0$ and $\gamma_3$, namely
\begin{eqnarray}
\tilde{\gamma}_0 &=& \frac{1}{\sqrt{\lambda_3^2 + \nu_3^2}}(\lambda_3\gamma_0 - \nu_3\gamma_3),\\
\tilde{\gamma}_3 &=& \frac{1}{\sqrt{\lambda_3^2 + \nu_3^2}}(\nu_3\gamma_0 + \lambda_3\gamma_3),
\end{eqnarray}
we can rewrite $H_\mathrm{eff}$ in a more suggestive form,
\begin{eqnarray}
H_\mathrm{eff}(\kappa) &=& 
 \frac{\sqrt{\lambda_3^2 + \nu_3^2}}{\lambda_3} (\lambda_3\kappa_3 - \nu\kappa_2)\,\tilde{\gamma}_3\nonumber\\
&&{}+ \frac{2k_{0,2}}{\lambda_3} (c\nu\kappa_3 + f\lambda_3\kappa_2)\,\gamma_2 \nonumber\\
&&{}+ \nu\kappa_1\gamma_1.
\end{eqnarray}
In addition, we define new coordinates $\tilde{\kappa}_2 = \lambda_3\kappa_3 - \nu\kappa_2$ and $\tilde{\kappa}_3 = c\nu\kappa_3 + f\lambda_3\kappa_2$, and collect the prefactors in new factors $v_i$. With that, the effective Hamiltonian becomes
\begin{equation}
H_\mathrm{eff}(\kappa) = 
v_1 \kappa_1 \gamma_1
+ v_2 \tilde{\kappa_2} \gamma_2
+ v_3 \tilde{\kappa_3} \tilde{\gamma}_3,
\end{equation}
which finally shows that $H_\mathrm{eff}$ has indeed the structure of a massless Dirac Hamiltonian.

\subsection*{B.2 Effective Hamiltonians for Weyl points and nodal lines}

By breaking inversion symmetry and thereby lowering the symmetry group of the $L'$ point from $C_{2h}$ to $C_s$, additional symmetry-allowed terms can be added to the Hamiltonian in Eq.~\eqref{eq:kp_Hamiltonian}. These terms are
\begin{eqnarray}
H_\mathrm{break} &=& \alpha\sigma_x + \beta(k_1 s_2 - k_2 s_1) 
+ \delta\beta(k_1 s_2 + k_2 s_1)\nonumber\\
&&{} + \xi(k_1 s_2 - k_2 s_1)\sigma_z
+ \delta\xi(k_1 s_2 + k_2 s_1)\sigma_z\nonumber\\
&&{} + \chi_1 k_1 s_3 \sigma_z + \chi_2 k_1 s_3
+ \chi_3 k_3 s_1\nonumber\\
&&{}+ \chi_4 k_3 s_1\sigma_z + \chi_5 s_1\sigma_y.
\end{eqnarray}  
For simplicity, in the main text we consider only the first two terms. We note, however, that all symmetry-allowed terms give rise to Weyl semimetal and nodal line phases, as can be explicitly checked numerically using the SnTe model.

Let us now add the term $\alpha\sigma_x$ to the $\mathbf{k}\cdot\mathbf{p}$ Hamiltonian of Eq.~\eqref{eq:kp_Hamiltonian}. The energies of the resulting Hamiltonian can be written as
\begin{eqnarray}
E^2 &=& (f_1 k_1^2 + f_2 k_2^2 + g k_2 k_3 + c k_3^2 + m)^2\nonumber\\ 
&&{}+ \Big[\alpha \pm \sqrt{k_1^2 (\lambda_1^2 + \nu_1^2) + (k_3 \lambda_3 - k_2 \nu_2)^2} \Big]^2\nonumber\\
&&{}+ (k_2 \lambda_2 + k_3 \nu_3)^2,
\end{eqnarray}
Since Weyl points will be located at $E=0$, their position in momentum space is determined by requiring that the \emph{three} binomials above are identical to zero. Hence, there are three polynomial equations for the three  momentum space coordinates $k_1,k_2,k_3$. We stress that, contrary to the existence condition for the Dirac points, there are no longer any conditions on the relation between the Hamiltonian parameters. 
This reflects the fact, that a Weyl point is a stable topological feature.

To simplify the discussion, we are now going to consider an effective Hamiltonian obtained by expanding around a Dirac point at $\mathbf{k}_0$. It reads
\begin{equation}
H_{\alpha,\mathrm{eff}}(\kappa) = H_\mathrm{eff}(\kappa) + \alpha\sigma_x,
\end{equation}
with the effective Dirac Hamiltonian $H_\mathrm{eff}$ from Eq.~\eqref{eq:effective_Dirac_Hamiltonian}.
If we instead add $\beta(k_1 s_2 - k_2 s_1)$, the effective Hamiltonian around $\mathbf{k}_0$ takes the form
\begin{eqnarray}
H_{\beta,\mathrm{eff}}(\kappa) &=& H_\mathrm{eff}(\kappa) 
+ \beta(s_1 \kappa_2 + s_2 \kappa_1 - k_{0,2} s_1).
\label{eq:effective_nodal_line_Hamiltonian}
\end{eqnarray}
Both effective Hamiltonians are used in the main text to derive approximate positions of Weyl nodes and nodal lines, respectively.

\section{C: Additional analysis of the nodal lines}

The zero-energy states of Hamiltonian $H_{\beta,\mathrm{eff}}$ from Eq.~\eqref{eq:effective_nodal_line_Hamiltonian} are located at
\begin{eqnarray}
\mathbf{k}_{N,\eta} &=& \frac{-\lambda_3\beta k_{0,2}}{\lambda_3\beta \pm \sqrt{A_\eta}}\,
(0,1,\eta),
\label{eq:nodal_lines_position_supp}
\end{eqnarray}
where 
\begin{equation}
A_\eta = 4\lambda_3^2 (f k_{0,2} + c k_{0,3}\eta)^2 
+ (\nu - \lambda_3\eta)^2 (\lambda_3^2 + \nu_3^2).
\end{equation}
In the following, we are going to show that the set of zero energy states parametrized by $\mathbf{k}_{N,\eta}$ forms a closed line topologically equivalent to a circle.

First of all, it is clear from the definition of $A_\eta$ that $A_\eta\geq 0$. Furthermore, $A_\eta$ is even strictly nonzero for all $\eta$ as we infer from solving $A_\eta=0$: the solutions are
\begin{eqnarray}
\eta_{1,2} &=& \frac{1}{\lambda_3(4c^2 k_{0,3}^2 + \lambda_3^2 + \nu_3^2)}
\bigg[ 
-4cf k_{0,2}^2 \nu + (\lambda_3^2 + \nu_3^2)\nu \nonumber\\
&&{} \pm 2\sqrt{-(f k_{0,2} \lambda_3 + c k_{0,3}\nu)^2 (\lambda_3^2 + \nu_3^2)} \bigg].
\end{eqnarray}
We immediately see that the term under the root is always negative. Hence, there are no real solutions of the equation $A_\eta=0$. Consequently, $A_\eta>0\: \forall\eta$ implies that the two solutions in Eq.~\eqref{eq:nodal_lines_position_supp} are always distinct, i.e. there are no crossings between the two branches of solutions.

Let us now look at Eq.~\eqref{eq:nodal_lines_position_supp} in the limits $\eta\rightarrow \pm\infty$. We obtain
\begin{eqnarray}
\lim_{\eta\rightarrow \pm\infty} \mathbf{k}_{N,\eta}^{(+)} &=& 
\pm \frac{\beta k_{0,2}}{\sqrt{4c^2 k_{0,3}^2 + \lambda_3^2 + \nu_3^2}}\,
(0,0,1),\\
\lim_{\eta\rightarrow \pm\infty} \mathbf{k}_{N,\eta}^{(-)} &=& 
\mp \frac{\beta k_{0,2}}{\sqrt{4c^2 k_{0,3}^2 + \lambda_3^2 + \nu_3^2}}\,
(0,0,1),
\end{eqnarray}
implying $\lim_{\eta\rightarrow \pm\infty} \mathbf{k}_{N,\eta}^{(+)} = \lim_{\eta\rightarrow \mp\infty} \mathbf{k}_{N,\eta}^{(-)}$. In other words, the two distinct solutions for the location of zero-energy states parametrized by $\eta$ are connected at $\eta=\pm\infty$. Hence, the solutions form indeed a closed line.

\end{document}